\documentclass[epj,twocolumn]{webofc}
\usepackage[varg]{txfonts}   

\newcommand{\be}{\begin{eqnarray}}
\newcommand{\ee}{\end{eqnarray}}

\woctitle{Advances in Dark Matter and Particle Physics
}
\begin{document}
\title{
Problems with the sources of the observed gravitational waves  and their resolution
}
%
%

\author{A.D. Dolgov  
\inst{1,2}\fnsep\thanks{\email{dolgov@fe.infn.it
    }} 
}

\institute{NSU, Novosibirsk, 630090, Russia
\and
 ITEP, Moscow, 117218, Russia
          }

\abstract{%
Recent direct registration of gravitational waves by LIGO and astronomical observations of the 
universe at  redshifts 5-10 demonstrate that the standard astrophysics and cosmology are in 
tension with the data. The origin of the source of the GW150914 event, which presumably is a 
binary of coalescing black holes with  masses about 30 solar masses, each with zero spin, as 
well as the densely populated universe at z= 5-10 by superheavy black holes, blight galaxies, 
supernovae, and dust does not fit the standard astrophysical picture. It is shown here that the 
model of primordial black hole (PBH) formation, suggested in 1993, nicely explains all these 
and more puzzles, including those in contemporary universe, such as MACHOs and the mass 
spectrum of the observed solar mass black holes.. The mass spectrum and density of PBH is 
predicted. The scenario may possibly lead to abundant antimatter in the universe and even in 
the Galaxy.
}
\maketitle
\section{Introduction}
\label{intro}
{The standard cosmological $\Lambda$CDM model (here CMD stands for Cold Dark Matter and $\Lambda$ for
vacuum-like dark energy, or, what is the same, $\Lambda$-term) 
very well describes gross features of the universe}  such as the spectrum of
perturbations at large scales, features of CMB (especially the shape of the angular fluctuation spectrum), 
baryogenesis, big bang nucleosynthesis, {\it etc} at expense of a few parameters.
However many ingredients of the standard cosmology are absent in 
the minimal standard model (MSM) of particle physics, in particular, dark matter, dark energy, baryogenesis, and
vacuum energy. To be more precise, the situation with vacuum energy is opposite: there is too much vacuum
energy in the MSM. For example the energy of the gluon and quark condensate established by quantum
chromodynamics are roughly 45 orders of magnitude higher than the observed magnitude of the dark energy.

So we have to conclude that new physics beyond the frameworks of MSM 
is a necessity. Still, except for the vacuum energy problem, the new physics may be almost the
old one with introduction of some new fields of particles, while it is quite possible that
for the solution of the vacuum energy problem a revolutionary modification of the
existing theory would be necessary.

In addition to these well known good old problems the
last several years revealed many features which look surprising  and even completely mysterious
in the frameworks of the $\Lambda$CDM model. These recent discoveries are reviewed in what follows.
It is argued that all them can be explained by formation of heavy primordial black holes (PBHs)
through the mechanisms suggested and discussed in detail in papers~\cite{pbh-form}. This mechanism
allows for formation of PBHs with the masses in the range from a fraction of the solar mass, say, up to
$10^4 M_\odot$ or even higher. Usually only the PBHs with rather low masses ${\sim 10^{20} }$ g 
were considered.

Moreover, the puzzling properties of the sources of the gravitational waves
recently discovered 
by LIGO~\cite{ligo-1} are explained by the same scenario of heavy PBH formation.

{The content of the talk is the following:} \\
{1. GW observation by LIGO.}\\
{2. Problems with the GW sources.}\\
{3. Solution of the problems and predictions.}\\
{4. Dense population of the universe at ${z\sim 10}$ by the objects which
could not be there.} They include in particularly {supermassive BHs,}
{early supernovae and gamma-bursters},
evolved chemistry and dust in high  ${z}$ universe. \\
{5. Problems in present day universe:} MACHOs, PBH dark matter, supermassive  BHs in 
large galaxies and even in almost empty space.

The talk is based on several our papers~\cite{bpp,beasts,bp-anti,ad-sb,bambi-ad}, where the relevant references can
be found.


\section{Direct discovery of gravitations waves} \label{s-GWdiscovery}

On February 11, LIGO (Laser Interferometer Gravitational wave Observatory) 
collaborations announced discovery of gravitational waves from a coalescing binary
systems of black holes~\cite{ligo-1}. Two more events were reported shortly, see below
Table.~\ref{fig-3events}.

The shape of the signal is in perfect agreement with the theory of BH interactions in 
the strong (Schwarzschild) sefl-fields, so it can be considered as a first direct proof of BH existence 
and a persuasive confirmation of General Relativity for a large deviation of geometry from the flat Minkowsky one.
All the previous tests were about weak fields only.

This discovery opened a new era of gravitational waves telescopes which will presumably allow to observe 
several (many) such catastrophic events per year. The anticipated increase of the LIGO sensitivity by factor
three will give enhance the number of the registered events 27-fold.
With the expected onset of operation of VIRGO (Italy) and
KAGRA (The Kamioka Gravitational Wave Detector, Japan) the direction to the sources can be 
reliably established, so the sources can be
studied by optical and other electromagnetic telescopes. New discoveries are imminent.

In Table I, copied from ref.~\cite{ligo-1}, the properties of the two coalescing black holes, 
which produced the powerful burst of gravitational waves,  are presented. 
The mass and spin of the final BH, and  the total energy radiated
in gravitational waves are estimated  by the fits to numerical simulations of binary black hole mergers.
{The estimated total energy radiated in gravitational waves is ${ (3.0 \pm 0.5) M_\odot }$ and the peak of 
gravitational-wave luminosity is ${3.0^{+0.5}_{-0.4} \times 10^{56} }$ erg/sec equivalent to 200${ M_\odot}$/sec,}
{more than whole radiation power of the visible universe.} 
Rotational energy  (outside the BH) is about ${0.3 M_\odot}$. It may be in principle extracted. 

\begin{figure}[htbp]
\centering
\includegraphics[width=9cm,clip]{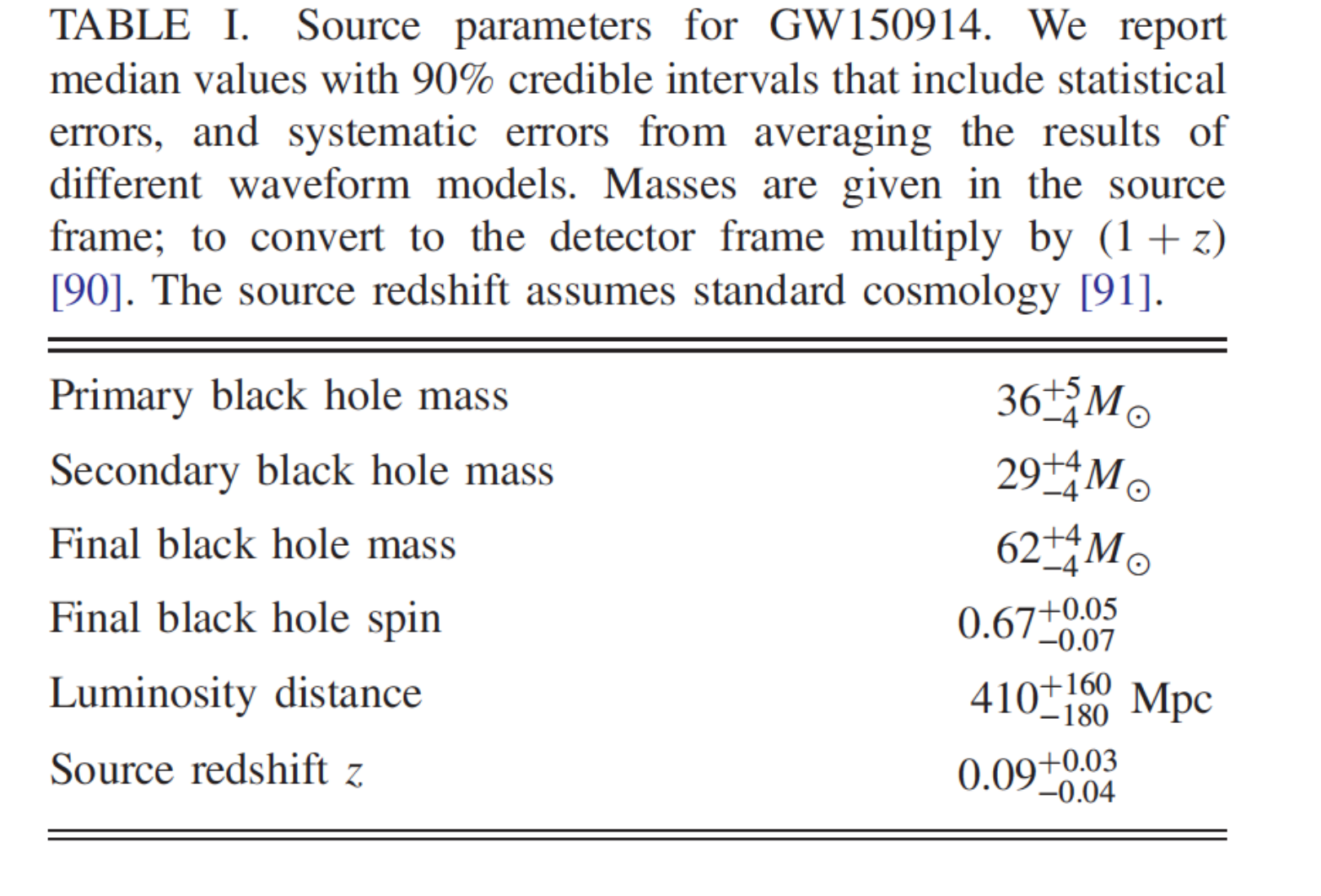}
\label{fig-GW1-results}       
\end{figure}

Three GW event observed to the present time are described in Fig.~\ref{fig-3events}.
The last event is questionable while the first two look reliable enough. The essential 
difference between GW150914 and GW151226 is the following: the masses of the sources in
the first case are very high, they are 
are 36 and 29 solar masses,  while in the second case they are
$14M_\odot$ and $7M_\odot$, which look astrophysically normal. The spins of both progenitors in 
the first case  are
low, compatible with zero. In the second case one of the black holes has noticeable spin $a>0.2$.
The formation mechanism of the initial back holes in the first case demands some unknown, unusual
astrophysical processes, while each of the companions on the second pair may be created through the usual
astrophysical channel. However, the formation of the binaries in both cases is not well understood, see below.

There are essentially  three problems in the standard theory:\\
{1. Origin of heavy BHs  (${\sim 30 M_\odot}$).}\\
{2. Low spins of the coalescing BHs.}\\
{3. Formation of BH binaries from original stellar binaries.}

{The first problem is a heavy BH origin.} 
{Such BHs are believed
to be created by massive star collapse,} though a convincing theory is still lacking.
{To form so heavy BHs, the progenitors should have ${M > 100 M_\odot}$
and  a low metal abundance to avoid too much
mass loss during the evolution.} Such heavy stars might be present in
young star-forming galaxies {but they are not yet observed in sufficiently high number.}

{ Another problem is the low value of the BH spins in GW150914.}
It strongly constrains astrophysical BH formation from close binary systems. {However, the 
dynamical formation of double massive low-spin BHs in dense stellar clusters is not excluded.}
{The second reliable LIGO detection, GW151226,
turned out to be closer to the standard binary BH system.}

{Last but not the least is the problem of formation of BH binaries.} Stellar binaries were 
formed from common interstellar gas clouds and are quite frequent in galaxies.
{If BH is created through stellar collapse,} {a small non-sphericity of collapse results in a huge 
velocity of the BH and the binary is destroyed.} An indirect evidence for that is
presented by large velocities of pulsars in the Galaxy. Their velocities are about 1000 km/sec,
while the average star velocities are only 200 - 300 km/sec.
Moreover, the BH formation from PopIII stars and subsequent formation of BH
binaries with 
${\sim(30+30) M_\odot}$ is analyzed in the literature and is
found to be negligible.

{All these problems are solved if the observed sources of GWs are the binaries of
primordial black holes (PBH).}\\
Here a model of PBH formation is presented which
{naturally reproduces the puzzling properties of GW150914,} 
{the rate of binary BH merging events inferred from the 
first LIGO science run,} 
{and provides seeds for early supermassive BH formation.}  These PBHs are created at rest
and so their mutual capture to form a binary is not inhibited after they loose their relative
velocity due e.g. to dynamical friction. The spins of the original BHs are naturally zero because 
rotational perturbations are absent in the early universe.

{In addition, the mechanism explains an avalanche of mysteries discovered  recently} 
and may provide all or a large fraction of cosmological DM in the form of PBHs with rather wide 
mass spectrum.

\section{Mechanism of massive PBH formation.} \label{PBH-form}

The model of an early BH creation is based on the supersymmetric
(Affleck-Dine) scenario for baryogenesis~\cite{affl-dine},
modified by introduction of a general renormalizable coupling to the inflaton 
field, see below, Eq.~(\ref{U-of-chi-Phi}).
It was suggested  in 1993~\cite{pbh-form}
and discussed in more details in several our papers applied to an
explanation of existence of the observed "old" objected in the young universe.
As a byproduct the model may lead to an abundant antimatter objects in the universe and,
in particular, in the Galaxy.

The basic ingredient of the Affleck-Dine (AD) scenario is a scalar field $\chi$ with
non-zero baryonic number, $ B\neq 0 $. As is well known, scalar baryons must exist in supersymmetric 
theories. Generically the potential of such field has the so called flat directions along which the potential
does not rise. We can take as toy model example the potential of the form:
\be
U_\lambda(\chi) = \lambda |\chi|^4 \left( 1- \cos 4\theta \right),
\label{U-of-chi}
\ee
where $\theta$ is the phase of the complex field $\chi = | \chi | \exp ( i \theta)$.
Such bosons, $\chi$, may condense along {flat} directions of this potential accumulating
large baryonic number. To be more precise, the baryonic number is accumulated 
not in the large amplitude of $\chi$ but in its angular momentum associated with variation of
$\theta (t)$, see eq. (\ref{B-chi}).

In addition to the quartic potential there may exist quadratic mass 
term, ${ m^2 \chi^2 + m^{*\,2}\chi^{*\,2}}$, which also could have flat directions generally different from
those in quartic poteetial: 
\be
U_m( \chi ) = m^2 |\chi|^2} {\left[{ 1-\cos (2\theta+2\alpha)} ,
\right],
\label{U-m-of-chi}
\ee
where ${ \chi = |\chi| \exp (i\theta)}$ and ${ m=|m|\,e^{ i \alpha}}$.
{If ${\alpha \neq 0}$, C and CP are  broken.}
{In GUT SUSY baryonic number is naturally non-conserved}. In out model this non-conservation
is induced by non-invariance of ${U(\chi)}$ w.r.t. the phase rotation of $\chi$. 

Initially (after inflation) ${\chi}$ was naturally away from the origin due to rising quantum 
fluctuations of light fields.  After 
inflation terminated,  $\chi$ started to evolve down to the equilibrium point, ${\chi =0}$,
according to the its equation of  motion, which for homogenous field formally coincides with the equation of motion
for point-like body in the Newtonian mechanics:
\be
\ddot \chi +3H\dot \chi +U' (\chi) = 0.
\label{ddot-chi}
\ee
The second term in this equation, induced by the universe expansion, is called the Hubble friction
term. In the mechanical analogy it is equivalent to liquid friction. 

The baryonic number of $\chi$ is equivalent to the angular momentum of $\chi$-rotation in two
dimensional complex plane $[Re\, \chi, Im\, \chi]$:
\be\
B_\chi =\dot\theta\, |\chi|^2 .
\label{B-chi}
\ee
Later the decays of ${{\chi}}$  transferred its
baryonic number to baryonic number of quarks in B-conserving process. The
Affleck-Dine baryogenesis could lead to the cosmological baryon asymmetry of order of unity, much larger
than the observed value $\beta=N_B/N_\gamma  \sim {10^{-9}}$.

When inflation terminated and the Hubble friction drastically dropped down, the field $\chi$ started to 
move to the origin, $\chi = 0$, along the flat direction of the quartic potential. At sufficiently small $\chi$
the quadratic mass term started to dominate and $\chi$ moved from the quartic valley to the quadratic
one gaining nonzero  and typically large angular momentum. This process occurred if the quartic and
quadratic flat directions are different. This surely happens if $\alpha \neq 0$ but may also happen even
with $\alpha = 0$ if initially $\chi$ was in a different quartic valley from that of the quadratic
$U_m( \chi ) $ (\ref{U-m-of-chi}).

{If the CP-odd phase ${\alpha}$ is small but non-vanishing, both baryonic and 
antibaryonic regions are possible }
{{with dominance of one of them.}
{Matter and antimatter domain may exist but globally ${ B\neq 0}$.}
 
In ref.~\cite{pbh-form} a general renormalizable coupling of $\chi$ to the inflaton field
$\Phi$ was introduced:
\be
\delta U ={g\, |\chi|^2 (\Phi -\Phi_1)^2} ,
\label{delta-U}
\ee
where $g$ is a dimensionless coupling constant and $\Phi_1$ is a value of the inflaton which it
passed during inflation. This potential looks as a very special one, but it is not so. It contains three 
renormalizable contributions: quartic, cubic, and quadratic terms. The only mild tuning of this potential
is the value of $\Phi_1$. 

With the new term and with an account of the so called Coleman-Weinberg contribution~\cite{CW},
arising from one-loop radiation corrections, the total potential governing the evolution of $\chi$ takes the form:
\be 
U& =& {g|\,\chi|^2 (\Phi -\Phi_1)^2}  +
\lambda |\chi|^4 \,\ln \left( \frac{|\chi|^2 }{\sigma^2 }\right)
\nonumber
\\
&+& \lambda_1 \left(\chi^4 + h.c.\right) + 
(m^2 \chi^2 + h.c.). \,\,\,\,\,\,\,\,\,\,\,\,\,\,\,\,\,\,\,\,
\label{U-of-chi-Phi}
\ee
The shape of this potential as a function of $| \chi |$ for different values of $\Phi$ is presented in Fig.~\ref{fig-phi-evol}.

When $\Phi$ is close to $\Phi_1$,
the window to the flat directions is open but only for a  relatively 
short period.  At this stage
cosmologically small but possibly astronomically large 
bubbles with high values of $\chi$ could be created. Later these high $\chi$ regions 
would create bubbles with very high baryonic number density
occupying {a small
fraction of the universe volume,
while the rest of the universe would have the normal baryon asymmetry
{${{ \beta \approx 6\cdot 10^{-10}}}$, created in the dominant part of the universe with small ${\chi}$}. 
The process when $\chi$ reaches large value but with low probability can be called
phase transition of 3/2 order, since in a sense such a behavior is between 
first and second order phase transitions.

The density contrast between the bubbles with high baryonic density (high B bubbles or HBB)
and the low B average cosmological background is initially small, since before the QCD phase 
transition quarks populating the cosmological plasma are essentially massless, so the 
initial density perturbations are predominantly isocurvature. However, after the phase transition to the
quark confinement phase quarks turned into heavy nucleons and the density contrast between HBB and
the rest of the universe became large.  Note that the contribution of HBBs into the total cosmological
energy/mass density can be higher than the contribution of the normal free baryons. 

The high density contrast, which appeared after the QCD phase transition, 
could lead to an early formation of compact stellar-type objects and
possibly to a comparable amount of anti-objects,
{such that the bulk of baryons and  (equal) antibaryons
are in the form of compact stellar-like objects or PBH,} 
{plus the sub-dominant observed homogeneous baryonic background.}
{The amount of antimatter may be comparable or even larger 
than of the known baryons,} 
{but such ``compact'' (anti)baryonic objects
do not contradict  any existing observations~\cite{bp-anti,bambi-ad}.

The distributions of high baryon density bubbles over length 
and mass have log-normal form~\cite{pbh-form}:
\be
\frac{dN}{dM} = C_M \exp{[-\gamma \ln^2 (M/M_0)]}
\label{dn-dM}
\ee
where ${C_M}$, ${\gamma}$, and ${M_0}$ are constant parameters.
{The spectrum is practically model independent, it is basically determined by inflation.}

\begin{figure*}
\centering
\includegraphics[scale=0.5]{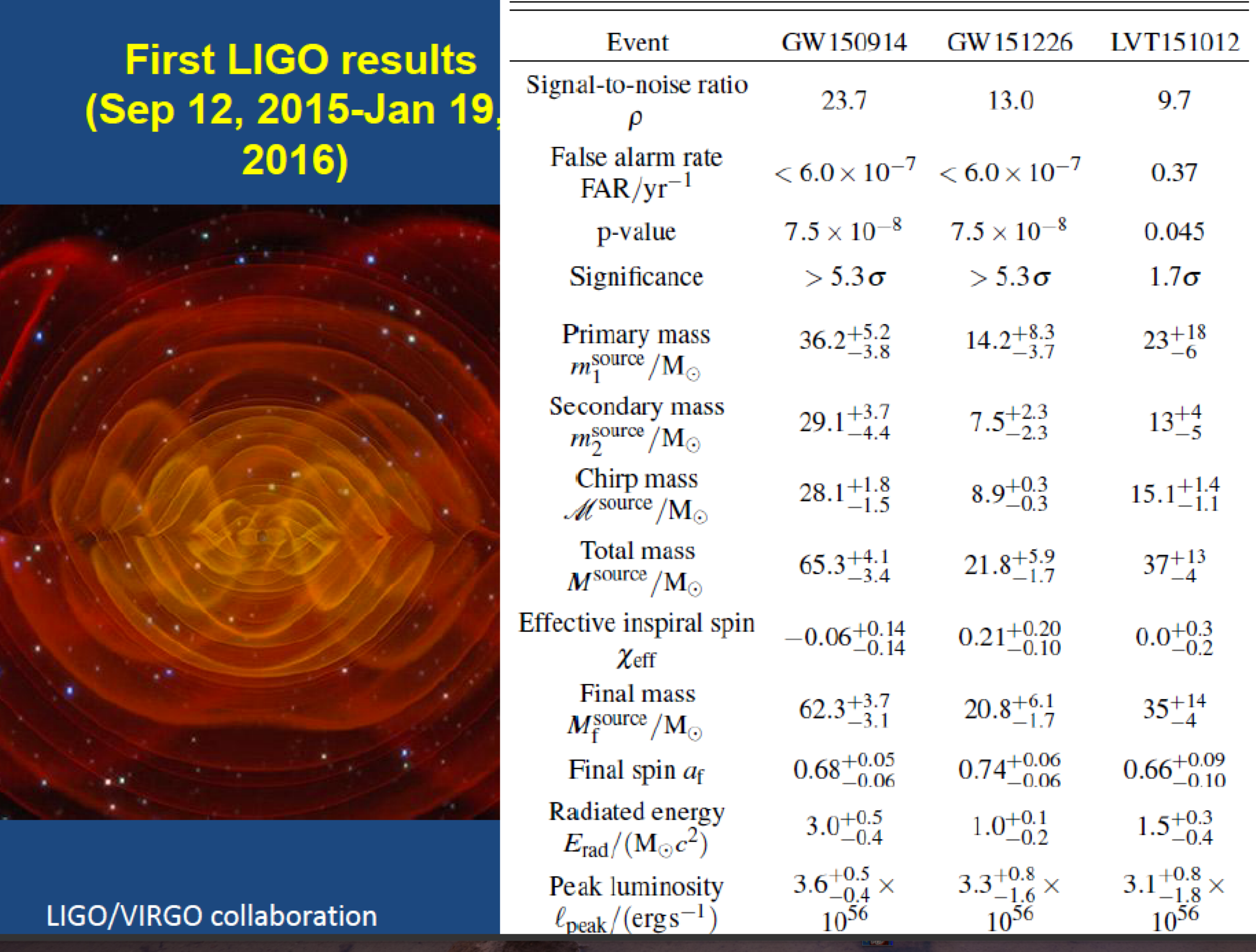}
\vspace*{5cm}       
\caption{Three registered gravitational wave events.}
\label{fig-3events}       
\end{figure*}

\begin{figure}[htbp]
\centering
\includegraphics[width=8cm,clip]{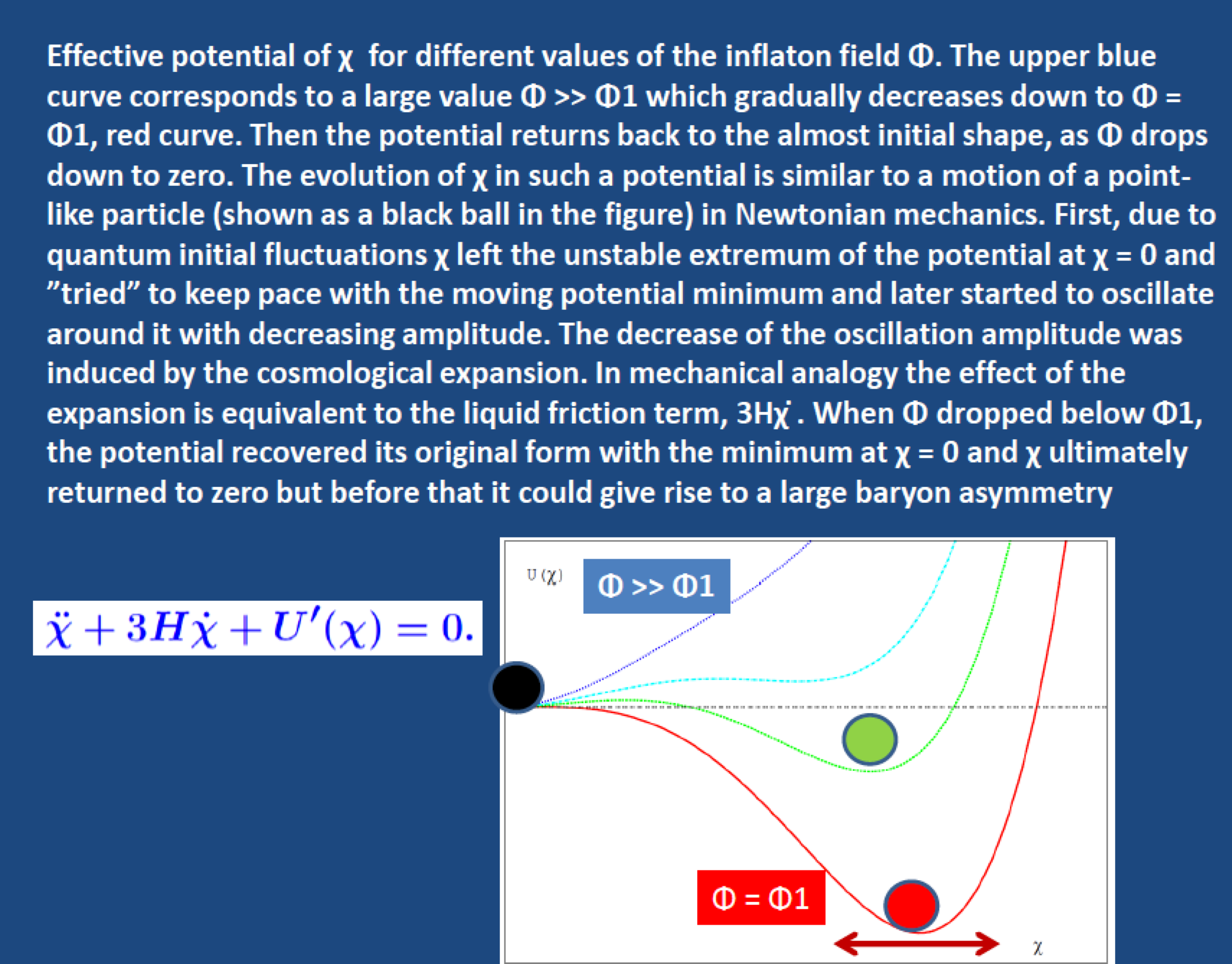}
\caption{Evolution of the potential (\ref{U-of-chi-Phi})   as a function of  $\chi $ for
different values of $\Phi$.}
\label{fig-phi-evol}       
\end{figure}

Adjusting parameters of the mass  spectrum (\ref{dn-dM}) using available astronomical data and 
constraints, we can predict the density of PBHs produced by the considered here mechanism. The
prediction as a function of the black hole mass together with the existing observational bounds
is presented in Fig.~\ref{fig-PBH}. 

\begin{figure}[htbp]
\centering
\includegraphics[width=9cm,clip]{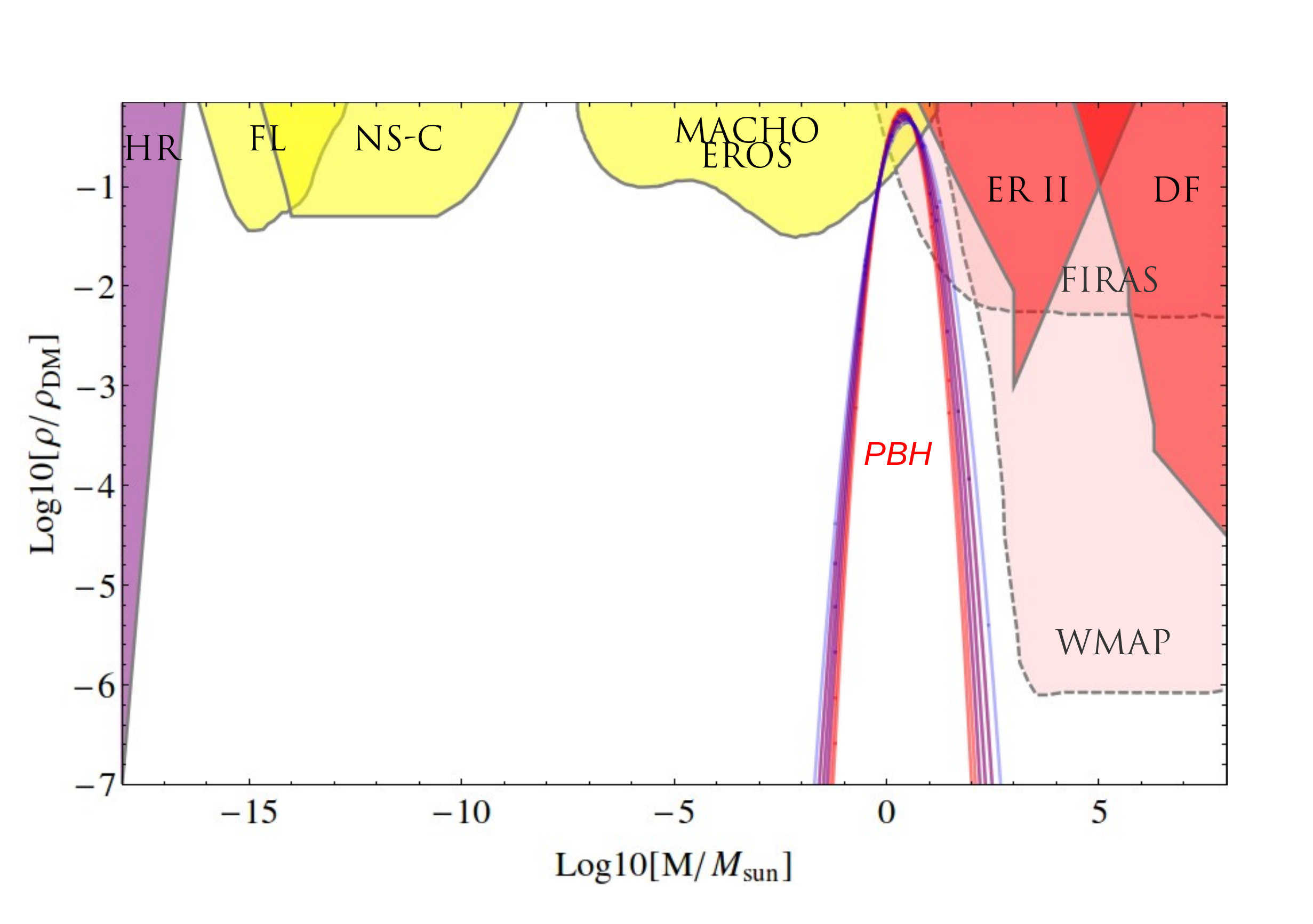}
\caption{Constraints on PBH fraction in DM, $f=\rho_{\rm PBH}/\rho_{\rm DM}$,
where the PBH mass distribution is taken as
$\rho_{\rm PBH}(M)=M^2dN/dM$. 
The existing constraints (extragalactic $\gamma$-rays from evaporation (HR),
femtolensing of $\gamma$-ray bursts (F),
neutron-star capture constraints (NS-C), MACHO, EROS, OGLE microlensing (MACHO, EROS)
survival of star cluster in Eridanus II (E),
dynamical friction on halo objects (DF),
and accretion effects (WMAP, FIRAS)).
The PBH distribution is shown for ADBD parameters $\mu=10^{-43}$~Mpc$^{-1}$,
$M_0=\gamma+0.1\times\gamma^2-0.2\times\gamma^3$ with $\gamma=0.75-1.1$ (red solid lines),
and $\gamma=0.6-0.9$ (blue solid lines).}
\label{fig-PBH}       
\end{figure}

\section{Problems in contemporary and near contemporary universe.} 
\label{sec-cont-univ}

\subsection{Dense  population of $z =5-10$ universe} \label{ss-hi-z}

Astronomical data accumulated during the last  few years have revealed  that the early, $ z \sim 10 $, 
universe is  unexpectedly dense, populated by the evolved objects which demand much more time
for their creation than was available at that high redshifts.
Among them there are bright but too young galaxies, 
{QSO/supermassive BHs,} {and  gamma-bursters (supernovae).} 
{Moreover, the early universe contains much more  dust than can be reasonably expected.} 
A more detailed review and the list of literature is presented in ref.~\cite{beasts}. Here we discuss only 
the most striking creatures.

About 40 quasars with ${z> 6}$ are already known, each quasar containing BH with
${M \sim 10^9 M_\odot}$. 
{Such black holes, {when the Universe was less than one billion years old,} 
present substantial challenges to theories of the formation and growth of
black holes and the coevolution of black holes and galaxies.}
{Even the origin of supermassive black holes in contemporary universe, which
had 14 Gyr  for their creation, is difficult to explain.}

Very recently a new monster was discovered
"An ultraluminous quasar with a twelve billion solar mass black hole at redshift 6.30"~\cite{QSO-10}.
{There is already a serious problem with formation of ten times lighter and less luminous quasars,}
{which is multifold deepened with this new "creature".}
A formation of the new one with ${M \approx 10^{10} M_\odot }$  is
absolutely impossible in the standard approach.

{The premature appearance of supermassive
black holes at ${z\sim 6}$} emerged as a great surprise. 
It is very difficult to understand how
{${10^9 M_\odot}$} black holes
appeared so quickly after the big
bang {without invoking non-standard accretion physics
and the formation of massive seeds,} 
both of which are not observed in the local Universe~\cite{melia}. On the other hand,
such massive seeds are exactly the objects which  are supplied by the HBBs discussed in the
previous section.

{Several galaxies have been observed at high redshifts,}
with natural gravitational lens ``telescopes. Their creation was also unexpectedly early.
E.g. a galaxy at {${z \approx 9.6}$} was discovered~\cite{gal-z9.6},
which was created when the universe was about 0.5 Gyr old.
Moreover a galaxy at {${z \approx 11}$} has been observed~\cite{gal-z11},
which was formed earlier than the universe age was { 0.41 Gyr} 
(or even shorter with larger H advocated by the recent traditional astronomical measurements).

An observation of not so young but extremely luminous galaxy is reported~\cite{elephant}. Its
luminosity is thousand times higher than that of the Milky way:
${L= 3\cdot 10^{14} L_\odot }$, while the age is only ${\sim 1.3 }$ Gyr.
The galactic seeds, or embryonic black holes, might be bigger than thought possible.
According to one of the authors of the discovery 
P. Eisenhardt: "How do you get an elephant?  One way is start with a baby elephant."
However, the origin of the baby elephant" is mysterious. 
The seed black hole should be already billions of solar masses, when our universe was only a 
tenth of its present age of 13.8 billion years. There is no way to create such a seed with 
standard mechanisms.

{The medium around the observed early quasars contains
considerable amount of ``metals''} (elements heavier than He). 
According to the standard picture, only elements up to ${^4}$He  { and traces of Li, Be, B}
were formed by the big bang nucleosynthesis, {while heavier elements were created much later
by stellar nucleosynthesis and} {dispersed in the interstellar space by supernova explosions.} It already
demands non-negligible time.

Much later molecular dust could form. Nevertheless it was discovered recently that 
the young universe at ${z >6}$ is quite dusty~\cite{dust}.
{Dusty galaxies show up at redshifts corresponding
to a Universe which is only about 500 Myr old.}
The highest redshift such object, HFLS3, lies at z=6.34 and
numerous other sources have been found~\cite{dust-2}.

{Hence, prior to or simultaneously with the QSO formation a rapid star formation should take place.}
{These stars should evolve to a large number of
supernovae enriching interstellar space by metals through their explosions}
which later make molecules and dust.
{(We all are dust from SN explosions, but probably at much later time.)}

Another possibility is a non-standard BBN due to very high baryonic density, which allows for 
formation of heavy elements beyond lithium.

Observations of high redshift gamma ray bursters (GBR) also indicate 
{a high abundance of supernova at large redshifts.} 
{The highest redshift of the observed GBR is 9.4} and there are a few more
GBRs with smaller but still high redshifts. 
{The necessary star formation rate for explanation of these early
GBRs is at odds with the canonical star formation theory.}

All such early supernovae can be HBBs which were not massive enough to make PBHs but
some compact stellar like objects which is also possible according to
the discussed in sec.~\ref{PBH-form} scenario.

\subsection{Back to the future.}\label{ss-back}

Similar and possibly related mysteries exist in contemporary and near-contemporary
universe.

An accumulation of quasars is discovered in a narrow spot in the sky~\cite{QSO-quartet}.
at redshift  ${z \approx 2}$.} According to the authors of the paper:
We discovered a physical association of four quasars embedded in a giant nebula. 
{Quasars are rare objects separated by cosmological distances, so
the chance of finding a quadruple quasar is ${\sim 10^{-7}}$.} 
It implies that the most massive structures in the distant universe have a tremendous supply 
(${\sim 10^{11} M_\odot}$) of cool dense (${ n \approx 1/}$cm${^3}$) gas,
{in conflict with current cosmological simulations."}

{Every large galaxy and some smaller 
ones contain central supermassive BHs} whose masses  are larger than 
{ ${ 10^{9}M_\odot}$} in giant elliptical
and compact lenticular galaxies
and {${\sim10^6 M_\odot}$} in spiral galaxies like Milky Way.
{The origin of these  superheavy BHs is not understood.}
{Moreover, SHBs  are observed  in several small galaxies,}
where is no material to make a supermassive BH.

It is intriguing if the type of the galaxy is determined by the mass of the original black hole seed.

The mass of BH is typically 0.1\% of the mass of the stellar bulge of galaxy
but some galaxies may  have huge BH: e.g. NGC 1277  has
the central BH
of  ${1.7 \times 10^{10} M_\odot}$, or ${60}$\% of its bulge mass~\cite{NGC1277}.
This fact creates serious problems for the
standard scenario of formation of central supermassive BHs by accretion of matter in the central 
part of a galaxy.

Report of the similar observations can be found in ref.~\cite{Khan}. The authors conclude that
although supermassive black holes
correlate well with their host
galaxies, there is an emerging view that outliers exist.
{ Henize 2-10, NGC 4889, and NGC1277 are examples of supermassive
BHs at least an order of magnitude more 
massive than their host galaxy suggests. }
{The dynamical effects of such ultramassive central black holes is unclear. }


Several more observations of too heavy black holes in poor galaxies are presented in ref.~\cite{beasts}
but the most striking one is given in the recent publication~\cite{naked}, where a supermassive BH was
observed in practically empty space

Thus the inverted picture is more plausible: first a supermassive black hole was formed and 
attracted matter serving as a seed for subsequent galaxy formation~\cite{pbh-form,inverted}.

\subsection{Old stars, black holes, and MACHOS in the Milky Way}\label{ss-old-stars}

{Some more, at first sight unrelated, but probably the same kind problems 
are demonstrated by the contemporary universe.}
{There are} {stars in the Milky Way, older than the Galaxy} and even 
{older than the universe} (more than two sigma) {and even one very old rocky planet.}
{The mass distribution of black holes in the Galaxy and abundant MACHOs are also  at odds with
the standard astrophysics.} 

The new recently developed precise measurements allowed to determine age of several stars in
the Milky Way with unprecedented accuracy. The results showed that quite a few stars are consuderably
older than it was earlier expected.

Employing thorium and uranium  abundances
in comparison with each other and with several stable elements {the age of
metal-poor, halo star BD+17$^o$ 3248 was estimated as}  ${13.8\pm 4}$ Gyr\cite{star-1}.
For comparison the age of inner halo of the Galaxy} ${11.4\pm 0.7}$ Gyr~\cite{inner-halo}.

The age of a star in the galactic halo, HE 1523-0901, was estimated to be 
about 13.2 Gyr~\cite{star-2}.
First time many different chronometers, such as the U/Th, U/Ir, Th/Eu, and Th/Os ratios to
measure the star age have been employed.

Even more striking, the metal deficient {high velocity} subgiant in the solar neighborhood
HD 140283  has the age ${14.46 \pm 0.31 }$ Gyr~\cite{star-3}.
The central value exceeds the universe age by two standard deviations,
if ${H= 67.3}$  km/sec/Mpc and ${t_U =13.8}$ Gyr. The excess is even bigger for
${H= 74}$ km/sec/Mpc, when ${ t_U = 12.5}$ Gyr.

A discovery of a surprisingly old planet was recently announced~\cite{planet}
Its age is estimated as  ${10.6^{+1.5}_{-1.3} }$ Gyr.
{(The age of the Earth: 4.54 Gyr.)} 
{A supenova explosion must  precede  formation of this planet.}

The considered in this report scenario of HBB formation may explain these striking discoveries, because 
if the initial chemical content of a star is different from the canonical one, it may
look older than it is in reality. 

Some more mysteries are revealed by the black holes observed in the Galaxy.
{It was found that the BH masses are concentrated in the narrow range
${ (7.8 \pm 1.2) M_\odot }$}~\cite{mass-bh-gal}.

 This result agrees with another paper where
a peak around ${8M_\odot}$, a paucity of sources with masses below
 ${5M_\odot}$, and a sharp drop-off above
${10M_\odot}$ are observed~\cite{mass-bh-gal-2}. 

{These features are not expected in the standard model, but may fit our model
 with the log-normal mass spectrum of PBHs.}

HBB can be also found by the effect of gravitational microlensing which may be caused by both
visible and invisible stars.
These objects are called Machos for Massive Astrophysical Compact Halo Objects. The observational
situation with them is reviewed in ref.~\cite{bp-anti}, where it is concluded  that
this population of the old invisible stars that evaded detection can well be HBBs
discussed here. Those stars should be
older than any kind of the oldest standard stellar populations. They will meet both criteria: they
should be very weak and their cloud should have such a high velocity dispersion as needed. There
is an intriguing possibility that some of those HBBs can be antistars in the Galaxy and in the galactic halo.

\section{Conclusion} \label{s-conclude}

{1. Supersymmetric baryogenesis could lead to abundant formation of PBHs and compact stellar-like
objects in the early universe after the QCD phase transition at $z\sim 10^{12}$ or
${t \gtrsim 10^{-5}} $ sec.}\\
{2. These objects have log-normal mass spectrum.} \\
{3. Adjusting the spectrum parameters is possible to explain the peculiar features of the sources
of the gravitational waves observed by LIGO.}\\
{4. The considered mechanism solves the numerous mysteries of ${z \sim 10}$ universe: abundant population
of supermassive black holes, 
early created gamma-bursters and supernovae, early bright galaxies, and evolved chemistry including dust}.\\
{5. There is persuasive data in favor of the inverted picture of galaxy formation, when first a supermassive BH seeds  are 
formed and later they accrete surrounding matter forming galaxies.}\\
{6. An existence of supermassive black holes observed  in all large and some small galaxies and even in
almost empty environment is naturally explained.}\\
{7. "Older than the universe" stars may exist.}\\
{8. Existence and high density of invisible "stars" (machos) can be understood.} \\
9. Some noticeable fraction of dark matter or even all of it can be made of PBHs.\\
10.  {Large amount of astronomical data the data strongly demand abundant cosmological 
population of PBH with wide mass spectrum.}
\\[3mm]
{Testable predictions:}\\
{A. Rate and masses of the BH sources of the coming  GW events.} \\
{B. Possible existence of antimatter in our neighborhood, even in the Galaxy.}

\section*{Acknowledgement}
This work was supported by the Grant of President of Russian Federation for
the leading scientific schools of the Russian Federation, NSh-9022.2016.2.

%
%
%

\end{document}